\documentclass[aps, pre, superscriptaddress]{revtex4-2} 
\usepackage{titlesec}

\titlespacing\section{0pt}{12pt plus 4pt minus 2pt}{8pt plus 2pt minus 2pt}

\usepackage{amsthm,amsmath}
\RequirePackage{hyperref}
\usepackage[utf8]{inputenc} 
\usepackage{nameref}

\def\includegraphics{}
\usepackage{graphicx}
\graphicspath{{./figures/}}

\usepackage{subcaption}
\usepackage{amsfonts}
\usepackage{dsfont}
\usepackage{xcolor}
\usepackage{lipsum}
\usepackage[nameinlink]{cleveref}

\crefname{equation}{Eq.}{Eqs.}
\Crefname{equation}{Equation}{Equations}
\crefname{paragraph}{Paragraph}{Paragraphs}

 \usepackage[multidot]{grffile}  

\newcommand{\be}{\begin{equation}}
\newcommand{\ee}{\end{equation}} 
\newcommand{\bea}{\begin{eqnarray}}
\newcommand{\eea}{\end{eqnarray}}

\newcommand{\grad}{{\nabla}} 

\newcommand{\rup}[1]{\left[#1\right]}

\usepackage[normalem]{ulem} 

\begin{document}
\title{Convergence properties of optimal transport-based temporal hypernetworks}
\author{Diego Baptista}
\affiliation{ Max Planck Institute for Intelligent Systems, Cyber Valley, Tuebingen, 72076, Germany}
\author{Caterina De Bacco}
\affiliation{ Max Planck Institute for Intelligent Systems, Cyber Valley, Tuebingen, 72076, Germany}

\begin{abstract}

We present a method to extract temporal hypergraphs from sequences of 2-dimensional  functions obtained as solutions to Optimal Transport problems.  We investigate optimality principles exhibited by these solutions from the point of view of hypergraph structures. Discrete properties follow patterns that differ from those characterizing their continuous counterparts. Analyzing these patterns can bring new insights into the studied transportation principles. We also compare these higher-order structures to their network counterparts in terms of standard graph properties. We give evidence that some transportation schemes might benefit from hypernetwork representations. We demonstrate our method on real data by analyzing the properties of hypernetworks extracted from images of real systems.
\end{abstract}

\maketitle 




\section*{Introduction}
Optimal Transport (OT) is a principled theory to compare probability distributions \cite{kantorovich1942transfer, villaniot, santambrogio2015optimal, peyre2019computational}. Although this task is usually framed as an  optimization problem, recent studies have mapped  it  within the framework of dynamic partial differential equations \cite{evans1999differential, facca2016towards, facca2019numerics, facca2021branching,tero2007mathematical, tero2010rules}. In this context, solutions to a transportation problem are often found as the convergent state of evolving families of functions. 
\\
In some scenarios, the steady states of these evolving families are supported in network-shaped structures \cite{xia2003optimal, xia2014landscape, Xia2015}. Recently, this fact has called the attention of network scientists and graph theorists leading to the development of methods that convert the solutions of OT problems into actual graph structures \cite{baptista2020network,leite2022revealing}. This has broadened the available set of tools to understand and solve these transportation problems. Recent studies have shown that common patterns can be unveiled in both the original mathematical setting and  in the converted graph structures \cite{baptista2021temporal}. 

Representations of these functions as sets of dyadic relations have been proven meaningful in various applications \cite{baptista2020principlednet,facca2021branching}.  Nonetheless, traditional dyadic representations may be limited in representing flows of quantities like mass or information as observed in real systems. Various examples of systems where interactions happen between 3 individuals or more are observed in applications as social contagion \cite{PhysRevResearch.2.023032,chowdhary2021simplicial}, random walks \cite{PhysRevE.101.022308,schaub2020random} or non-linear consensus \cite{PhysRevE.101.032310}. Understanding the relation between the structure and
 dynamics taking place on higher-order structures is an active field of research \cite{taylor2015topological,patania2017topological}. For instance, key elements controlling dynamics are linked to the heterogeneity of hyperedges' sizes present in their higher-order representations \cite{patania2017topological}. These systems are hence best described by hypergraphs, generalizations of networks that encode structured relations among any number of individuals. With this in mind, a natural question to ask is how do OT-based structures perform in terms of higher-order representations?
\\
To help bridge this knowledge gap about higher-order properties of structures derived from OT solutions, we elaborate on the results observed in \cite{baptista2021temporal}. Specifically, we propose a method to convert the families of 2-dimensional functions into temporal hypernetworks. We enrich the existing network structures associated with these functions by encoding the  observed interactions into hyperedges. We study classic hypergraph properties and compare them to the predefined cost functional linked to the transportation problems.  Finally, we extend this method and the analysis to study systems coming from real data. We build hypergraph representations of \textit{P. polycephalum} \cite{westendorf2016quantitative} and analyze their topological features.

\section*{Methods}\label{section:methods}
\subsection*{The Dynamical Monge-Kantorovich Method}
\paragraph*{The Dynamical Monge-Kantorovich set of equations.} We start by reviewing the basic elements of the mechanism chosen to solve the OT problems. As opposed to other standard optimization methods used to solve this \cite{cuturi2013sinkhorn}, we use an approach that turns the problem into a dynamical set of partial differential equations. In this way, initial conditions are updated until a convergent state is reached. The dynamical system of equations as proposed by Facca et al. \cite{facca2016towards,facca2019numerics,facca2021branching}, is presented as follows.  We assume that the OT problem is set on a continuous 2-dimensional space $\Omega  \in \mathbb{R}^{2}$, and at the beginning,  no  underlying network structure is observed. This gives us the freedom of exploring the whole space to design an optimal network topology, solution of the transportation problem. The main quantities that need to be specified in input are \textit{source} and \textit{target} distributions. We refer to them as sources and sinks, where a certain mass (e.g. passengers in a transportation network, water in a water distribution network) is injected and then extracted. We denote these with a ``forcing'' function $f(x)=f^+(x)-f^-(x)\in \mathbb{R}$,  describing the flow-generating sources $f^+(x)$ and sinks $f^-(x)$. To ensure mass balance it is imposed $\int_\Omega f(x)dx  = 0$. We assume that the flow is governed by a transient Fick-Poiseuille  flux $q=- \mu \grad u$, where $\mu,u$ and $q$ are called \textit{conductivity} (or \textit{transport density}), \textit{transport potential} and \textit{flux}, respectively. Intuitively, mass is injected through the source, moved based on the conductivity across space, and then extracted through the sink. The way mass moves determines a flux that depends on the pressure exerted on the different points in space; this pressure is described by a potential function.

The set of \textit{Dynamical Monge-Kantorovich} (DMK) equations is given by:
\begin{align}
-\nabla \cdot (\mu(t,x)\nabla u(t,x)) &= f^+(x)-f^-(x) \,, \label{eqn:ddmk1}\\
\frac{\partial \mu(t,x)}{\partial t}  &= \rup{\mu(t,x)\nabla u(t,x)}^{\beta} - \mu(t,x) \,, \label{eqn:ddmk2}\\
\mu(0,x)  &= \mu_0(x) > 0  \label{eqn:ddmk3} \,,
\end{align}
where $\nabla=\nabla_{x}$. \Cref{eqn:ddmk1} states the spatial balance of the Fick-Poiseuille flux and is complemented by no-flow Neumann boundary conditions. \Cref{eqn:ddmk2} enforces the dynamics of this system, and it is controlled by the so-called \textit{traffic rate} $\beta$. It determines the transportation scheme, and it shapes the topology of the solution: for $\beta<1$ we have congested transportation where traffic is minimized, whereas  $\beta>1$ induces branched transportation where traffic is consolidated into a smaller amount of space. The case $\beta=1$ recovers shortest path-like structures. Finally, \Cref{eqn:ddmk3} constitutes the initialization of the system and can be thought of as an initial guess of the solution.   

Solutions $(\mu^*, u^*)$ of \crefrange{eqn:ddmk1}{eqn:ddmk3} minimize the transportation cost function $\mathcal{L}(\mu,u)$ \cite{facca2016towards,facca2019numerics,facca2021branching}, defined as:
\begin{align}\label{eqn:L}
& \mathcal{L}(\mu,u) := \mathcal{E}(\mu,u)+ \mathcal{M}(\mu,u) \\
& \mathcal{E}(\mu,u) := \dfrac{1}{2}\int_{\Omega} \mu |\grad u|^2 dx, \  \ \mathcal{M}(\mu,u) :=  \dfrac{1}{2}\int_{\Omega} \dfrac{\mu^{\frac{(2-\beta)}{\beta}}}{2-\beta} dx \quad.
\end{align}

$\mathcal{L}$ can be thought of as a combination of $\mathcal{M}$, the total energy dissipated during  transport (or network operating cost) and $\mathcal{E}$, the cost to build the network infrastructure (or infrastructural cost). It is known that this functional's convexity changes as a function of $\beta$. Non-convex cases arise in the branched schemes, inducing fractal-like structures \cite{facca2021branching,  santambrogio2007optimal}. This is the case that we considered in this work, and it is the only one where meaningful network  structures, and thus, hypergraphs, can be extracted \cite{baptista2020network}. 

\subsection*{Hypergraph sequences}
\paragraph*{Hypergraph construction.} We define a hypergraph (also, hypernetwork) as follows \cite{battiston2020networks}: a \textit{hypergraph} is a tuple $H = (V, E),$ where $V = \{v_1, ... ,v_n\}$ is the set of \textit{vertices} and $E = \{ e_1, e_2, ... , e_m\}$ is the set of \textit{hyperedges} in which $e_i\subset V, \forall i = 1,...,m,$ and $|e_i|>1$.   If $|e_i|=2,\forall i$ then $H$ is simply a graph.  We call \textit{edges} those hyperedges $e_i$ with $|e_i|=2$ and \textit{triangles}, those with $|e_i|=3$.   We refer to the \textit{1-skeleton} of $H$ as the \textit{clique expansion} of $H$. This is the graph $G=(V,E_{G})$ made of the vertices $V$ of $H$, and of the pairwise edges built considering all the possible combinations of pairs that can be built from each set of nodes defining each hyperedge in $E$.

Let $\mu$ be the conductivity found as a solution of \crefrange{eqn:ddmk1}{eqn:ddmk3}. As previously mentioned, $\mu$  at convergence regulates where the mass should travel for optimal transportation. Similar  to \cite{baptista2021temporal}, we turn this 2-dimensional function into a different data structure, namely, a hypergraph.  This is done as follows: consider $G(\mu) = (V_G,E_G)$ the network extracted using the method proposed in \cite{baptista2020network}. We define $H(\mu)$ as the tuple $(V_H,E_H)$ where $V_H = V_G$ and $E_H = E_G \cup T_G,$ s.t., $T_G = \{(u,v,w):  (u,v),(v,w),(w,u) \in E_G, \}.$ In words, $H(\mu)$ is the graph $G(\mu)$ together with all of its triangles. 
This choice is motivated by the fact that the graph-extraction method proposed in \cite{baptista2020network} uses triangles to discretize the continuous space $\Omega$, which can have a relevant impact on the extracted graph or hypergraph structures. Hence, triangles are the natural sub-structure for hypergraph constructions. The method proposed in this work is valid for higher-order structures beyond triangles. Exploring how these additional structures impact the properties of the resulting hypergraphs is left for future work.  

\begin{figure}[!ht]
    \centering
\begin{subfigure}[b]{0.95\textwidth}
\includegraphics[width=\textwidth]{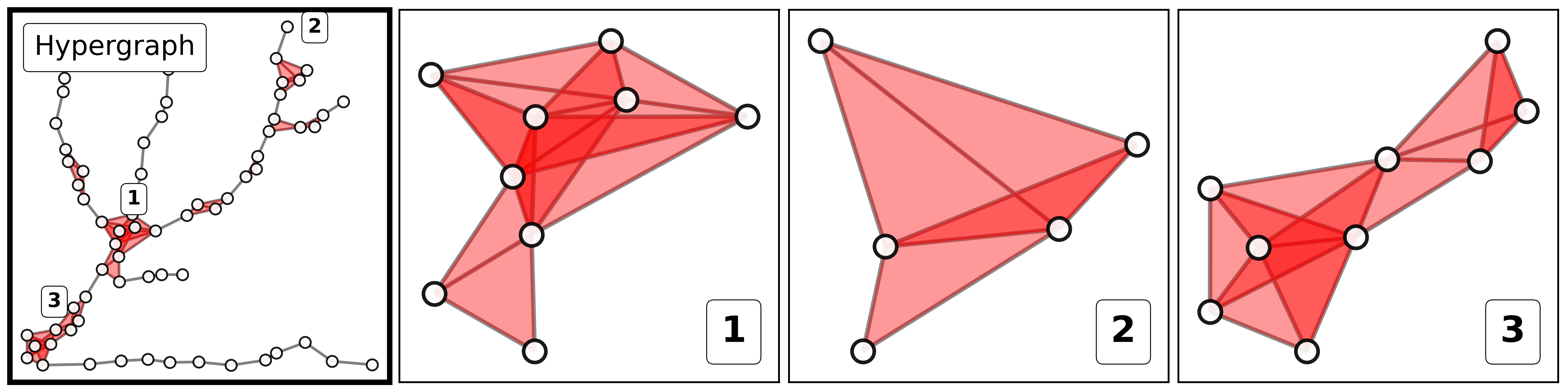}
\end{subfigure} 
\caption{\textbf{Hypernetwork construction.} Higher order structures are built using edges and triangles as hyperedges. The leftmost panel shows one of the studied graphs together with the triangles (in red) used. The subsequent panels highlight different clusters of triangles that can be seen in the main hypergraph.} \label{fig:image1}
\end{figure}

\Cref{fig:image1} shows an example of one of the studied hypergraphs. The red shapes represent the different triangles of $H(\mu)$. Notice that, although we consider here the case where $|e|\leq 3$ for each hyperedge $e$---for the sake of simplicity---higher-order structures are also well represented by the union of these elements, as shown in the right panels of the figure.

Since this hypergraph construction method is valid for any 2-dimensional transport density, we can extract a hypergraph not only from the convergent $\mu$ but also at any time step before convergence. This then allows us to represent optimal transport sequences as hypergraphs evolving in time, i.e. temporal hypernetworks. 

\begin{figure}[!ht]
    \centering
\begin{subfigure}[b]{0.9\textwidth}
\includegraphics[width=\textwidth]{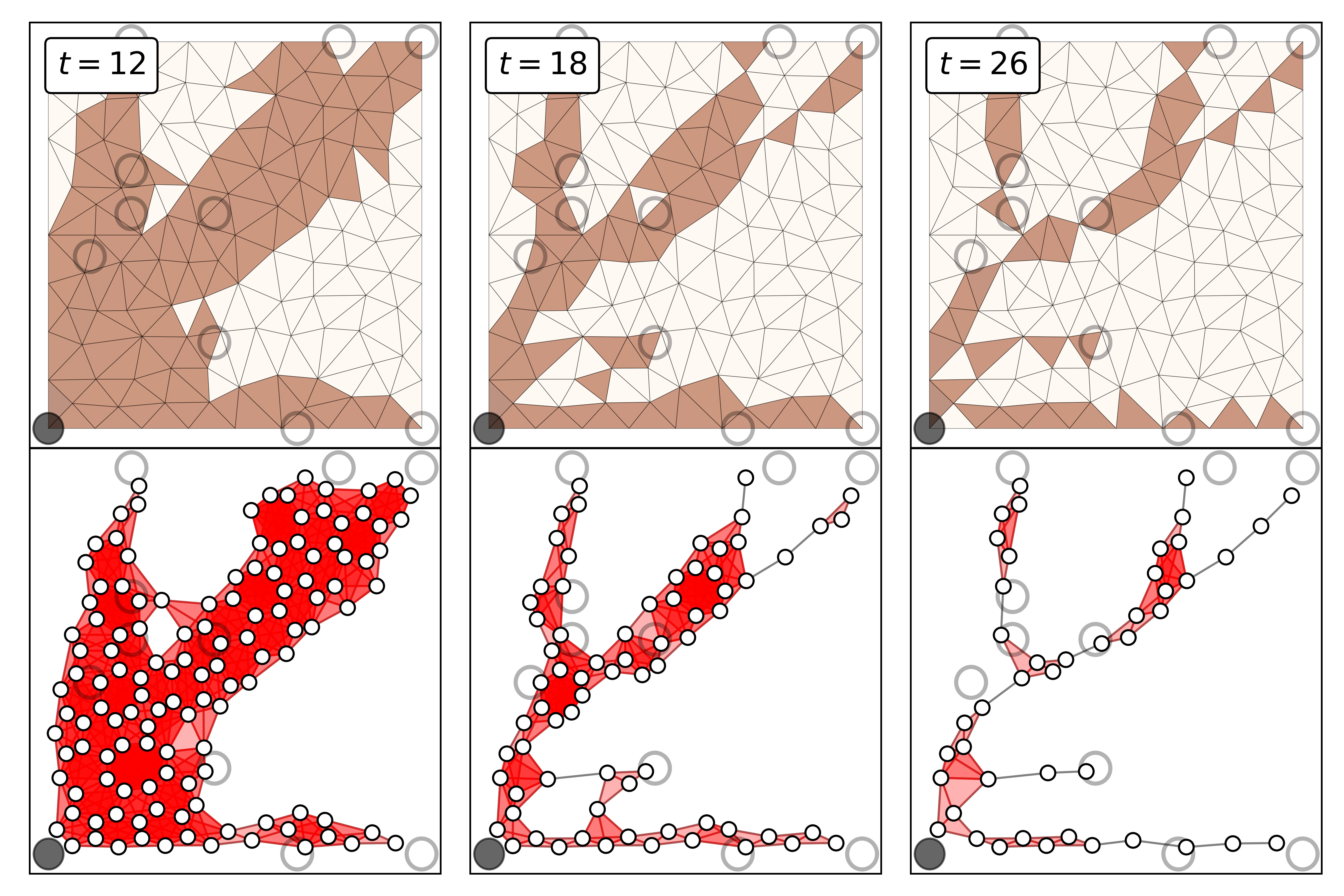}
\end{subfigure} 
\caption{\textbf{Temporal hypergraphs.}  Top row: different timestamps of the sequence $\{\mu_t\}$; triangles are a  discretization of $[0,1]^2$. Bottom row: hypergraphs extracted for $\mu_t$ at the time steps displayed on the top row; triangles are highlighted in red. In both rows, filled and empty circles correspond to the support of $f^+$ and $f^-$, i.e. sources and sinks, respectively. This sequence is obtained for $\beta = 1.5$.} \label{fig:image2}
\end{figure}

\paragraph*{Hypergraph sequences.} Formally, let $\mu(x,t)$ be a \textit{transport density} (or \textit{conductivity}) function of both time and space obtained as a solution of the DMK model. We denote it as the sequence $\{\mu_t\}_{t=0}^T$, for some index $T$ (usually taken to be that of the convergent state). Each $\mu_{t}$ is the $t$-th update of our initial guess $\mu_0$, computed by following the rules described in \crefrange{eqn:ddmk1}{eqn:ddmk3}. This determines a sequence of hypernetworks $\{ H(\mu_t)\}_{t=0}^T$ extracted from $\{\mu_t\}_{t=0}^T$ with the extraction method proposed in \cite{baptista2020network}. \Cref{fig:image2} shows three hypergraphs built from one of the studied sequences $\{\mu_t\}$ using this method at different time steps. The corresponding OT problem is that defined by the (filled and empty) circles: mass is injected in the bottom left circle and must be extracted at the highlighted destinations. On the top row, different updates (namely, $t=12, 18, 26$) of the solution are shown. They are defined on a discretization of $[0,1]^2.$ Darkest colors represent their support. Hypergraphs extracted from these functions are displayed at the bottom row. As can be seen, only edges (in gray) and triangles (in red) are considered as part of $H(\mu_t)$. Notice that the larger the $t$ is, the less dense the hypergraphs are, which is expected for a uniform initial distribution $\mu_0$ and branched OT ($\beta>1$) \cite{facca2021branching}.

\subsection*{Graph and hypergraph properties}
We compare hypergraph sequences to their correspoding network counterparts (defined as described in the previous paragraph). We analyze the following main network  and hypergraph properties for the different elements in the sequences and for different sequences. Denote with $G = (V_G,E_G)$ and $H = (V_H, E_H)$ one of the studied graphs and hypergraphs belonging to some sequence $\{ G(\mu_t)\}_{t=0}^T$ and $\{ H(\mu_t)\}_{t=0}^T$, respectively. We consider the following network properties:
\\
\begin{enumerate}
\item $|E_G|$, total number of edges;
\item Average degree $d(G)$, the mean number of neighbors per node;
\item Average closeness centrality $c(G)$: let $v\in V_G$, the closeness centrality of $v$ is defined as $
\sum_{u\in V_G} 1/d(u,v),$ where $d(u,v)$ is the shortest path distance between $u$ and $v$. 
\end{enumerate}

\begin{figure}[!ht]
    \centering
\begin{subfigure}[b]{\textwidth}
\includegraphics[width=0.98\textwidth]{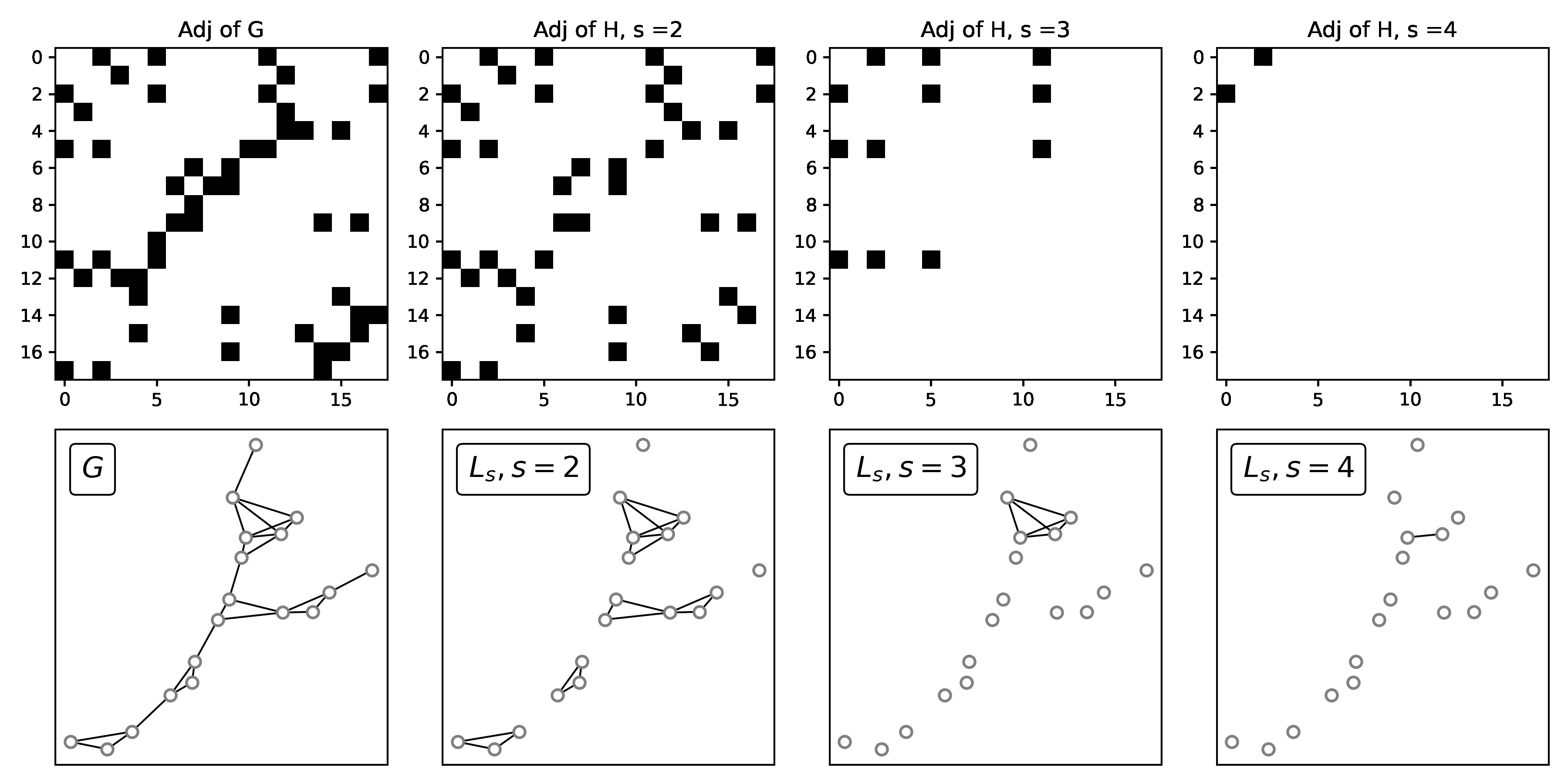}
\end{subfigure} 
\caption{\textbf{Adjacency matrices and line graphs.} Top: generalized node $s$-adjacency matrices for different values of $s$ from a given toy graph $G$. Bottom, from left to right: reference network $G$, and $s$-line graphs for $s=2,3,$ and $4$. } \label{fig:image3}
\end{figure}

Hypernetwork properties can be easily adapted from the previous definitions with the help of generalized adjacency matrices and line graphs \cite{aksoy2020hypernetwork}.  Let $H$ be a hypergraph with vertex set $V = \{1,..,n\}$ and edge set $E = \{e_1, ... ,e_m\}$.  We define the generalized \textit{node} $s$-\textit{adjacency matrix} $A_s$ of $H$ as the binary matrix of size $n\times n$, s.t., $A_s[i][j]=1$ if $i$ and $j$ are part of at least $s$ shared hyperedges; $A_s[i][j]=0,$ otherwise. We define the $s$-\textit{line graph} $L_s$ as the graph generated by the adjacency matrix $A_s$.  Notice that $A_1$ corresponds to the adjacency matrix of $H$'s skeleton (which is $L_1$). \Cref{fig:image3} shows a family of adjacency matrices together with the line graphs generated using them.  We can then define hypergraphs properties in the following way:
\\
\begin{enumerate}
\item $|E_H|$, total number of hyperedges;
\item $|T| = |\{e \in E_H: |e|= 3\}|,$ total number of triangles;
\item $S = \sum_{t\in T} a(t),$ \textit{covered area}, where $a(t)$ is the area of the triangle $t;$  
\item Average degree $d_s(H)$, the mean number of incident hyperedges of size greater or equal than $s$ per node; 
\item Average closeness centrality $c_s(H)$: let $v\in V_H$, the closeness centrality of $v$ is defined as its closeness centrality in $L_s$.
\end{enumerate}

\begin{figure}[!ht]
    \centering
\includegraphics[width=0.98\textwidth]{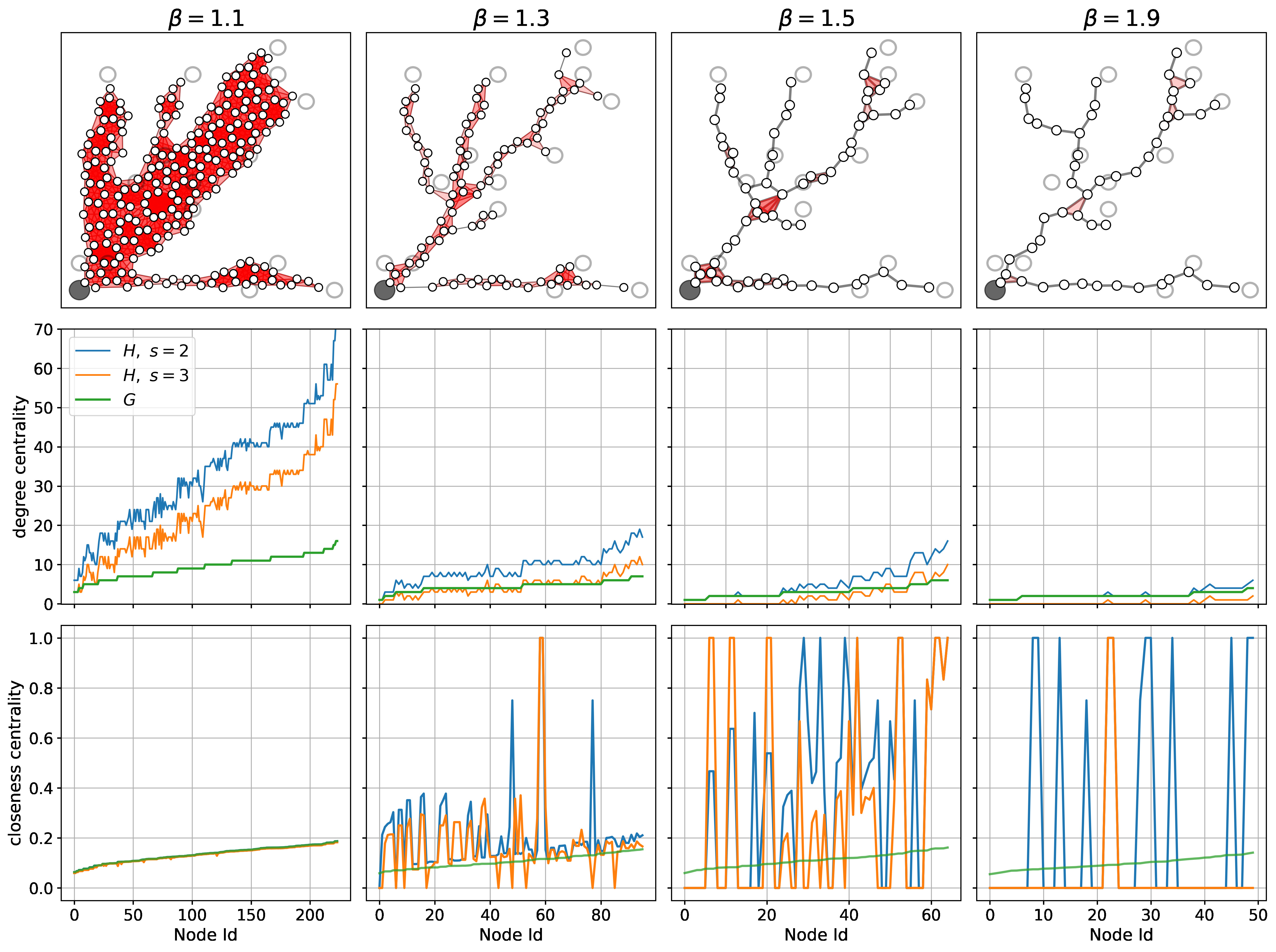}
\caption{\textbf{Graph and Hypergraph properties.}  Top row: optimal hypernetworks obtained with different traffic rates. Center and bottom rows: degree distributions and closeness distributions for the hypernetworks shown on the top row, and their 1-skeletons. The node labels in the $x$-axis of the center and bottom rows are sorted by their degree of centrality values.} \label{fig:image4}
\end{figure}

$S$ can be defined in terms of any other property of a hyperedge, e.g. a function of its size $|e|$. Here we consider the area covered by a hyperedge to keep a geometrical perspective.  On the other hand, this area $S$ can be easily generalized to hyperedges with $|e_{i}|>3$ by suitably changing the set $T$ in the summation, e.g. by considering structures containing four nodes.  As for the centrality measures, we focus our attention to compare the case $s>1$ against $s=1$, as the latter traces back to standard graph properties and we are interested instead to investigate what properties are inherent to hypergraps. \Cref{fig:image4} shows values of the $d_s(H)$ and $c_s(H)$ for convergent hypergraphs $H$ (obtained from different values of $\beta$) together with the degree and closeness centrality of their correspondent graph versions. The considered hypergraphs are displayed in the top row of the figure.   As can be seen in the figure, patterns differ considerably for different values of $\beta$. As $s$ controls the minimum number of shared connections for different nodes in the networks, the higher this number, the more restrictive this condition becomes, thus leading to more disconnected line graphs. In the case of the $s$-degree centrality, we observe decreasing values for increasing $s$, with nodes with the highest centrality having much higher values than nodes less central. For both $s=2,3$ we observe higher values than nodes in $G$. This follows from the fact that once hyperedges are added to $G$, the number of incidences per node can only increase.  Centrality distributions strongly depend on $\beta$. For small values---more distributed traffic ($\beta=1.1$)---the number of hyperedges per node remains larger than the number of regular edges connected to it. But if traffic is consolidated on less space ($\beta=1.9$), then very few hyperedges are found. This suggests that the information learned from hypergraphs that is distinct to that contained in the graph skeleton is influenced by the chosen traffic regime.

As for the closeness centrality distribution, this resembles that of $G$  for small values of $\beta$, regardless $s$. For higher $\beta$ it switches towards an almost binary signal. Thus, nodes tend to become more central as $\beta$ increases, suggesting that adding hyperedges to networks $G$ leads to shorter distances between nodes.  The loss of information seen for the highest values of $s$ is due to the fact that the line graphs $L_s$ become disconnected with many small connected components. In these cases, the closeness centrality of a node is either 0 if it is isolated, or proportional to the diameter of the small connected component where it lives in.

\paragraph{Convergence criteria.} Numerical convergence of the DMK \crefrange{eqn:ddmk1}{eqn:ddmk3}  is usually defined by fixing a threshold $\tau$. The updates are considered enough once the cost associated to them does not change more ($\leq \tau$) than that of the previous time step. As it is usually the case when this threshold is too small ($\tau=10^{-12}$ in our experiments), the cost or the network structure may consolidate to a constant value earlier than algorithmic convergence. Similar to \cite{baptista2021temporal}, to meaningfully establish when is hypergraph optimality reached, we consider as convergence time the first time step when the transport cost, or a given network property, reaches a value that is smaller or equal to a certain fraction $p$ of the value reached by the same quantity at algorithmic convergence (in the experiments here we use $p=1.05$). We refer to $t_\mathcal{L}$ and  $t_P$ for the convergence in times in terms of cost function or a network property, respectively.

\section*{Results}
To test the properties presented in the previous section and understand their connection to transportation optimality, we synthetically generate a set of optimal transport problems, determined by the configuration of sources and sinks.  As done in \cite{baptista2021temporal}, we fix a source's location and sample several points in the set $[0,1]^2$ to be used as sinks' locations. Let  $S = \{s_0,s_1,...,s_M\}$ be the set of locations in the space $[0,1]^2,$ and fix a positive number $0<r$. We define the distributions $f^+$ and $f^-$ as $
f^+(x) \propto  \mathds{1}_{R_0}(x),$ and   $f^-(x) \propto  \sum_{i>0} \mathds{1}_{R_i}(x),$  where $\mathds{1}_{R_i}(x) := 1,$ if $x\in R_i$, and $\mathds{1}_{R_i}(x) := 0$, otherwise; $R_i = C(s_i,r)$ is the circle of center $s_i$ and radius $r$. The value of $r$ is chosen based on the used discretization, and as mentioned before, the centers are sampled uniformly at random. The   symbol $\propto$ stands for proportionality and is used to ensure that $f^+$ and $f^-$ are both probability distributions.  The transportation cost is that of \cref{eqn:L}.

\paragraph{Synthetic OT problems.}\label{sec:synthetic}
The set of transportation problems considered in our experiments consists of 100 source-sink configurations. We place the location of the source $s_0=(0,0)$ (i.e. the support of $f^+$ at $(0,0)$), and sample 15 points $s_1,s_2,...,s_M$  uniformly at random from a regular grid. By sampling them from the nodes of the grid, we ensure that two different locations are at a safe distance so they are considered different once the space is discretized.  We initialize $\mu_0(x)=1, \forall x$ to be a uniform distribution on $[0,1]^2$. This can be interpreted as a non-informative initial guess for the solution. Starting from $\mu_0,$ we compute a maximum of 300 updates. Depending on the chosen traffic rate $\beta$ more or fewer iterations can be needed.  We claim that the sequence $\{\mu_t\}_{t=0}^T$ \textit{converges} to a certain function $\mu^*$ at iteration $T$ if either $|\mu_T-\mu_{T-1} |<\tau,$  for a \textit{tolerance} $\tau\in (0,1],$ or $T$ reaches the mentioned maximum.  For the experiments reported in this manuscript, the tolerance $\tau$ is set to be $10^{-12}$. Given the dependence of the solution of traffic constraints, a wide range of values of $\beta$ is considered. Namely, we study solutions obtained from low traffic cases ($\beta=1.1$, and thus, less traffic penalization) to large ones ($\beta=1.9$), all of them generating branched transportation schemes. Our 100 problems are linked to a total of 900 hypergraph sequences, each of them containing between 50 and 80 higher-order structures. 

\begin{figure}[!h]
    \centering
\includegraphics[width=\textwidth]{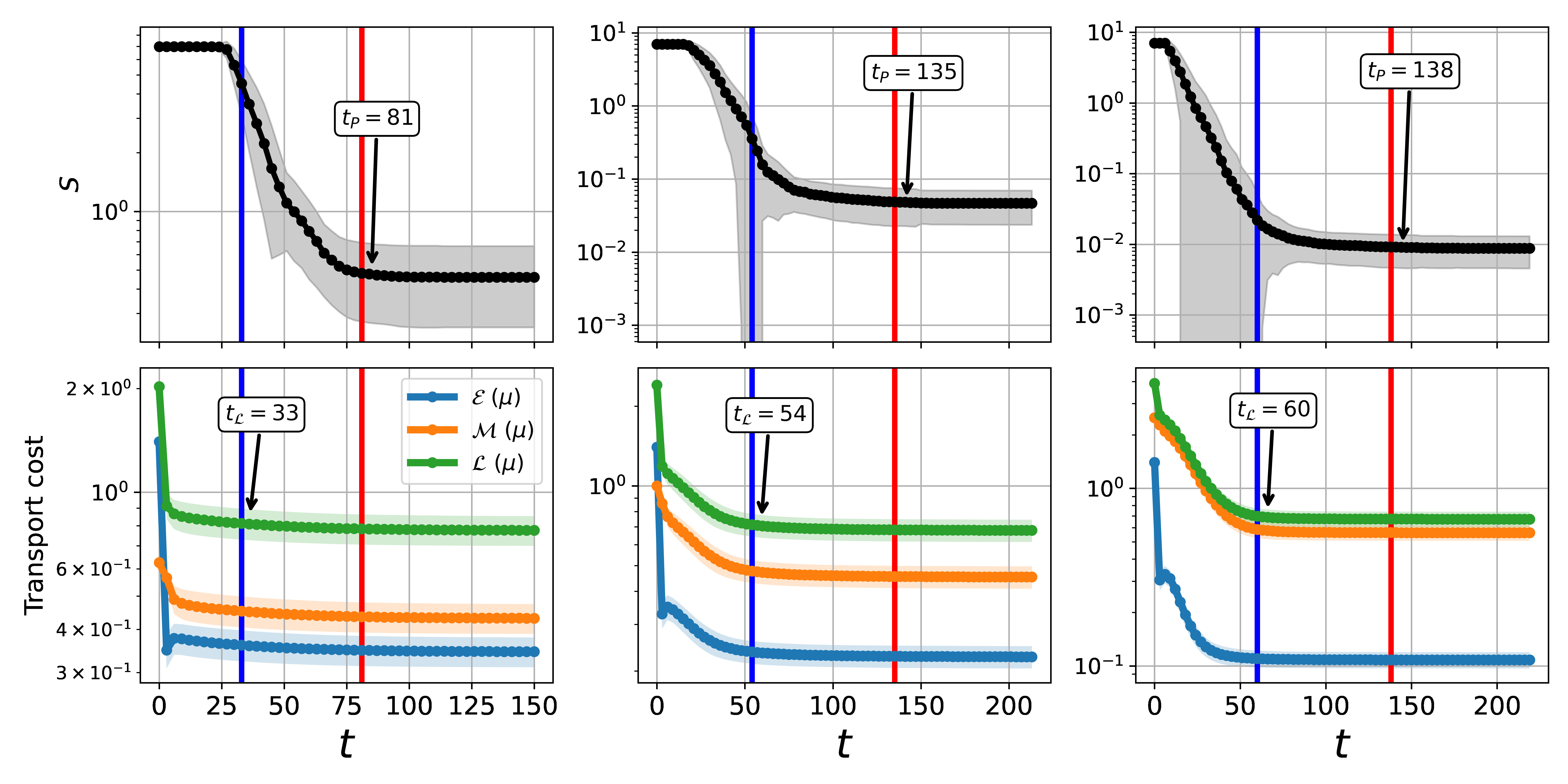}
\caption{\textbf{Covered area and Lyapunov cost.} Mean (markers) and standard deviations (shades around the markers) of the covered area $S$ (top plots) and  of the Lyapunov cost, energy dissipation $\mathcal{E}$ and structural cost $\mathcal{M}$ (bottom plots), as functions of time $t$. Means and standard deviations are computed on the set described in Paragraph \textit{Synthetic OT problems}. From left to right: $\beta=1.2, 1.5$ and $1.8$. Red and blue lines denote $t_P$ and $t_\mathcal{L}$.}  \label{fig:image5}
\end{figure}

\paragraph{Convergence: transport cost vs hypernetwork properties.}  

As presented in \cite{baptista2021temporal}, we show a comparison between hypernetwork properties and the cost function minimized by the dynamics, where convergence times are highlighted (\Cref{fig:image5}). We focus on the property $S$, the area of the surface covered by the triangles in $H$. This quantity is influenced by both the amount of triangles (hence of hyperedges) and their distribution in space. Hence, it is a good proxy for how hypergraph properties change both in terms of iteration time and as we tune $\beta$.
We observe that $t_P>t_\mathcal{L}$ in all the cases, i.e. convergence in terms of transportation cost is reached earlier than the convergence of the topological property.  Similar behaviors are seen for other values of $\beta\in[1.1,1.9]$ and  other network properties (see Appendix). Similar to DMK-based network properties, the covered area's decay is faster for the smallest values of $\beta$. This is expected, given the convexity properties of $\mathcal{L}$ \cite{facca2016towards,facca2019numerics,facca2021branching}. However, the transport cost decays even faster, in a way that the value of $S$ is still far away from convergence in the congested transportation case (small $\beta$).
\\
 Notice that  $S$ remains stable after the first few iterations, and then it starts decreasing at different rates (depending on $\beta$) until reaching the converged value. This suggests that the dynamics tend to develop thick branches---covering a large area--- at the beginning of the evolution, and then it slowly compresses them until reaching the optimal topologies. 
\\
These different convergence rates for $S$ and $\mathcal{L}$ may prevent construction of converged hypernetwork topologies: if the solver is stopped at $t_\mathcal{L}< t_{P}$, the resulting hypergraphs $H(\mu_t), \ t=t_\mathcal{L}$ would mistakenly cover a surface larger than that covered by the convergent counterpart ($H(\mu_t),$ for $t\geq t_P$). This scenario is less impactful for larger values of $\beta$, although in these scenarios $H$ is much more similar to a regular graph, because of the small number of higher-order structures. Topological differences between converged hypernetworks can be seen in \Cref{fig:image4}.
\\
Finally, we observe that both $t_\mathcal{L}(\beta)$ and $t_P(\beta)$ are increasing functions on $\beta$. This is expected since the larger the traffic rate is, the longer it takes for the sequences to converge. This particular behavior matches what is shown in \cite{baptista2021temporal} in the case of $t_\mathcal{L}$, but this is not the case for $t_P(\beta)$:  it was observed a non-monotonic behavior in the network case.

\begin{figure}[!h]
    \centering
        \includegraphics[width=0.8\textwidth]{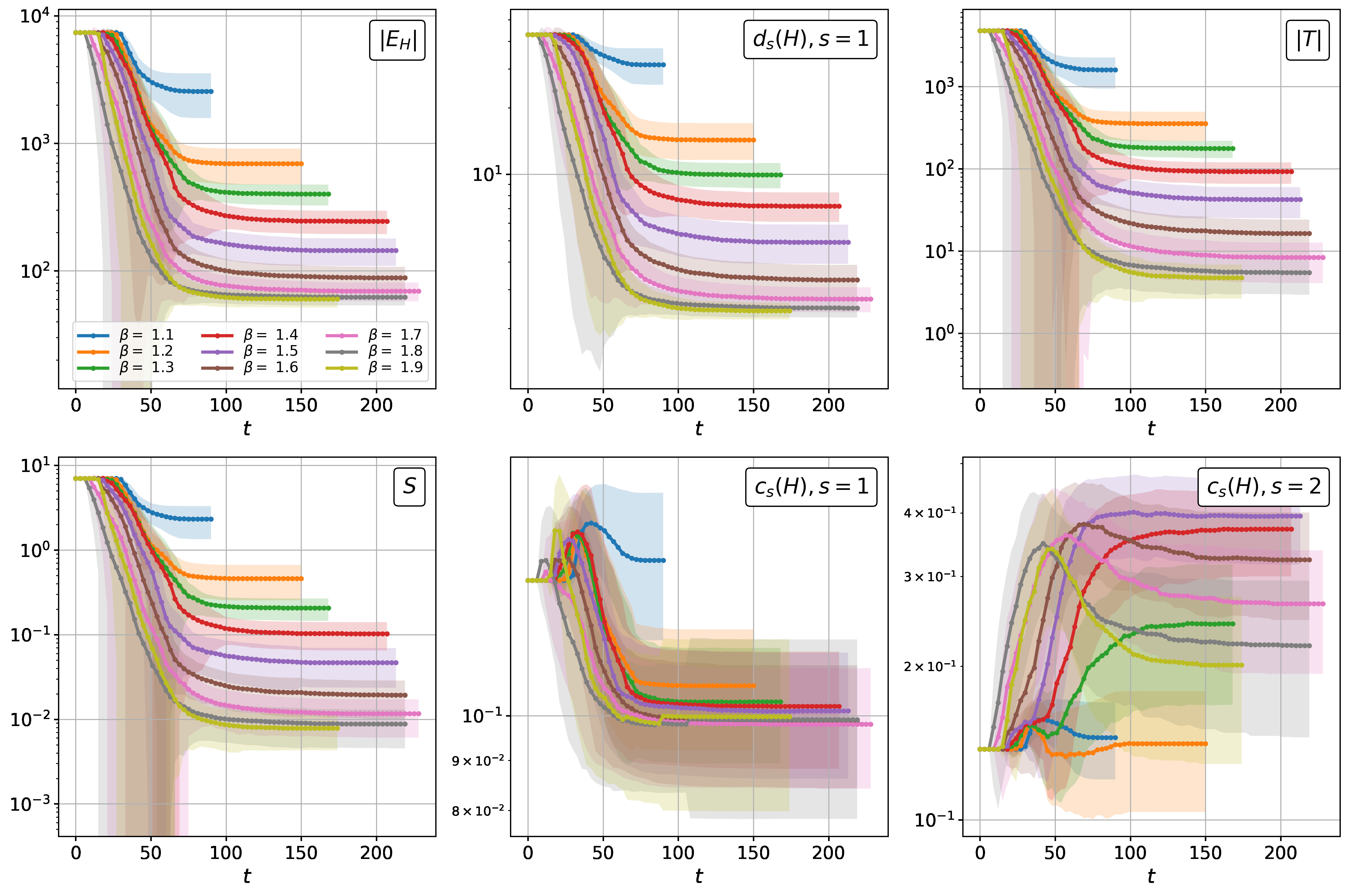}      
    \caption{\textbf{Evolution of hypernetwork properties}. Mean (markers) and standard deviations (shades around the markers) of number of hyperedges $|E_H|$ (upper left), number of triangles  $|T|$ (upper center), covered area $S(H)$ (upper right), average $2$-degree $d_2(H)$ (lower left), average $1$-closeness centrality $c_1(H)$(lower center) and $2$-closeness centrality $c_2(H)$(lower right), computed for different values of $\beta$ as a function of time.}\label{fig:image6}
\end{figure}

\paragraph{Convergence behavior of hypernetwork properties.}  
\Cref{fig:image6} shows how the various network properties change depending on the traffic rate. Mean values and standard deviations are computed across times, for a fixed value of $\beta$. As shown, the number of hyperedges, number of triangles, covered area, and average 1-degree exhibit decreasing patterns as functions of $t$. As a consequence, transport optimality can be thought of as reaching minimum states on the mentioned hypernetwork properties. Another clear feature of these functions is related to the actual converged values: the larger the $\beta$ is, the smaller these metrics become. This is explained by a cost function increasingly encouraging consolidations of paths on fewer edges. Notice also that the gap between these converged values signals a non-linear dependence on the outputs of the dynamics; e.g., a converged hypernetwork obtained for $\beta=1.1.$ loses many more hyperedges if the traffic rate is then set to 1.2, whereas this loss would not be that large if $\beta=1.2$ is increased to 1.3. The nature of these gaps is substantially different depending on the property itself. This also shows that certain properties better reveal  the distinction between different optimal traffic regimes. 


The behavior of the closeness centralities is distinctly different than that of the other properties. While its initial values are the same for all values of $\beta$ (similar to the previous properties), no clear trend can be found as time increases. For  $s=1$, on average $\beta=1.1$ generates sequences that tend to recover initial values after increasing and then decreasing behavior. For the other traffic rates, we observe different patterns. Notice that $s-$closeness centrality on the hypergraph for $s=1$ is the same as the classic closeness centrality on the skeleton of it. Thus, these rather noisy patterns are not due to the addition of hyperedges. On the other hand, for $s=2$ the average centrality  shows increasing curves. This may be due to $L_s$ getting increasingly disconnected with small connected components. Therefore, the larger $s$, the closer the nodes are seen (see \Cref{fig:image3}).  Moreover, in this case small values of $\beta$ lead to more stable closeness centrality values, showing the impact of $\beta$ in  building higher-order structures. While different values of $\beta$ lead to different behaviors of the hypergraph properties (e.g. decreasing degrees and amount of hyperedges for increasing $\beta$) we remark that choosing the value of $\beta$ should depend on the application at hand. The analysis performed here showcases how this choice may impact the resulting topologies. This can help practitioners to visualize possible consequences in terms of downstream analysis on the transportation properties of the underlying infrastructure.  

\section*{\textit{P. polycephalum} hypernetworks}

We now analyze hypernetworks extracted from images of real data. We are interested in the evolution of the area covered by triangles in the sequences $ \{ H(\mu_t)\}_{t=0}^T$ extracted from real images of the slime mold \textit{P. polycephalum}. The behavior of this organism is the inspiration of the modeling ideas of the DMK equations described in \nameref{section:methods}. It has been shown that these slime molds follow a similar optimization strategy as that captured by the DMK dynamics while foraging for food in 2D surfaces \cite{nakagaki2000maze,tero2007mathematical,tero2010rules}. 
  We extract hypernetworks from images using the idea described in  \nameref{section:methods} but instead of applying \cite{baptista2020network} to obtain the networks, we use the method proposed by \cite{baptista2020principlednet} which takes images as input. This pipeline uses the color intensities of the different image pixels to build a graph, by connecting adjacent meaningful nodes. We dedicate our attention to 4 image sequences from the Slime Mold Graph Repository \cite{dirnberger2017introducing}. The sequences are then describing the evolution of a \textit{P. polycephalum} placed in a rectangular Petri dish. Each image, and thus each hypernetwork, is a snapshot of the movement of this organism over periods of 120 seconds. 

We study the covered area for every one of the 4 sequences, and plot the results for one of them (namely, image set \textit{motion12}; see Appendix) in \Cref{fig:image7}. We highlight 4 times along the property sequence to display the used images together with the corresponding hypernetworks. The lower leftmost plot shows a subsection of one of the studied snapshots. As can be seen there, this subhypernetwork topology exhibits a significant number of hyperedges of dimension 3, mainly around the thickest parts of the slime mold. On the other side, in the lower rightmost plot,  the evolution of $S$ is overall decreasing in time (similar results are obtained for other sequences,  as shown in the Appendix). This suggests that the thicker body parts tend to get thinner as the \textit{P. polycephalum} evolves into a consolidated state.  This pattern resembles what is shown above for the synthetic data, i.e. the covered area tends to decrease as time evolves similar to the behavior of the DMK-based hypernetwork sequence.  This suggests that the DMK model realistically mirrors a consolidation phase towards optimality of real slime molds \cite{dirnberger2017introducing}. 

\begin{figure}[!ht]
    \centering
\begin{subfigure}[b]{1\textwidth}
\includegraphics[width=0.95\textwidth]{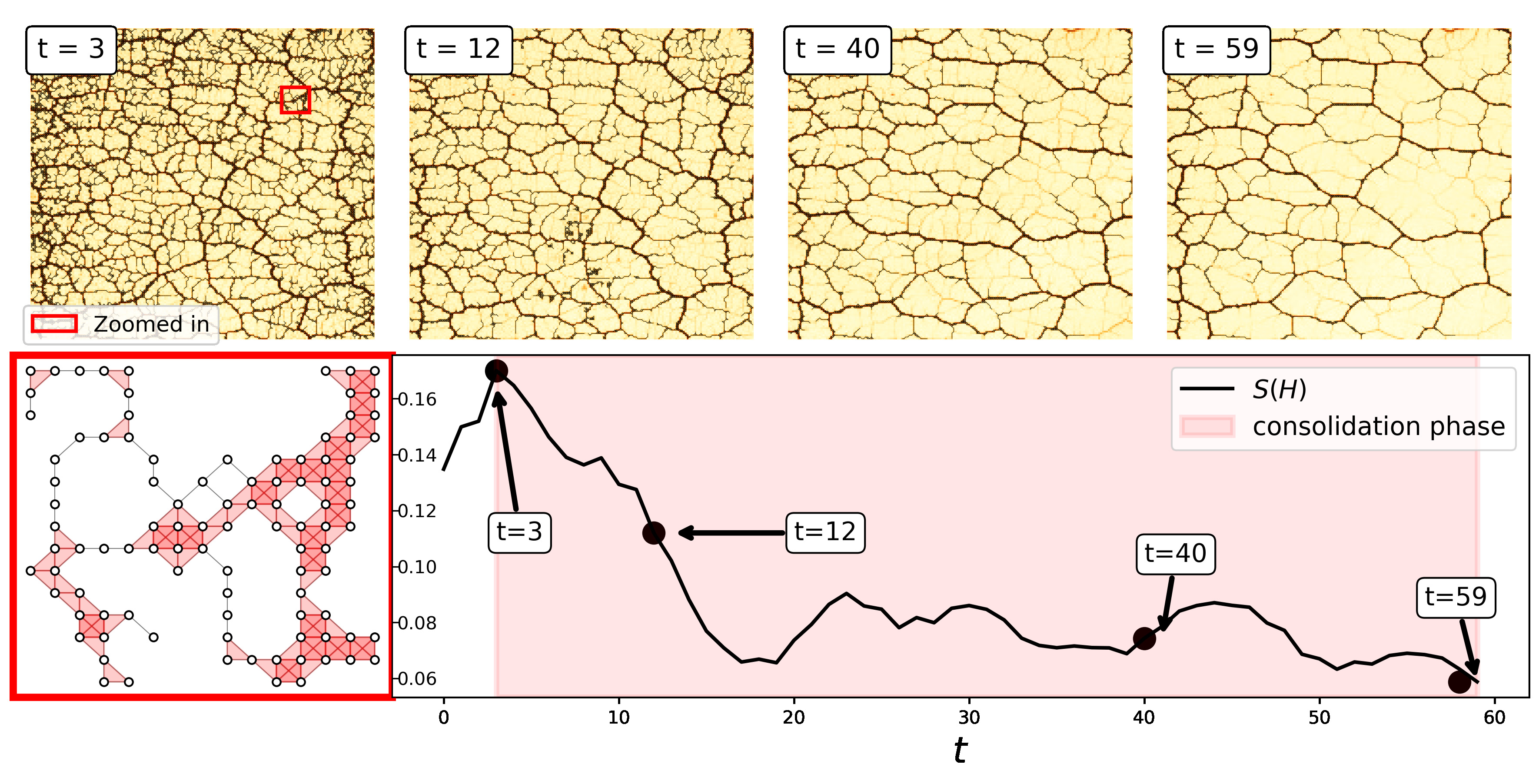}
\end{subfigure} 
\caption{\textbf{\textit{P. polycephalum} hypergraphs.} On top: \textit{P. polycephalum} images and hypernetworks extracted from them. Bottom left: a zoomed-in part of the hypergraph shown inside the red rectangle on top. Bottom right: covered area as a function of time. The red shade highlights a tentative consolidation phase towards optimality.} \label{fig:image7}
\end{figure}

\section*{Conclusions}
We proposed a method to build higher-order structures from OT sequences. This method maps every member of the sequence into a hypergraph, outputting a temporal hypernetwork. We analyzed standard hypergraph properties on these temporal families and compared them to their continuous counterparts. We showed that convergence in terms of transportation cost tends to happen faster than that given by the covered area of the hypernetworks. This suggests that the dynamics used to solve the OT problems concentrates the displaced mass into main branches, and once this task is carried out, it slightly reduces the area covered by them. We studied this and other hypergraph properties, and compared them to those of their network versions. In some cases, hypernetworks reveal more information about the topology at convergence. This suggests that hypernetworks could be a better alternative representation to solutions of OT problems for some transportation schemes. The conclusions found in this work may further enhance our comprehension of OT solutions and the links between this field and that of hypergraphs.

\paragraph{Acknowledgements}
The authors thank the International Max Planck Research School for Intelligent Systems (IMPRS-IS) for supporting Diego Baptista.


\bibliographystyle{splncs03} 
\bibliography{references} 

\newcommand{\beginsupplement}{%
        \setcounter{table}{0}
        \renewcommand{\thetable}{S\arabic{table}}%
        \setcounter{figure}{0}
        \renewcommand{\thefigure}{S\arabic{figure}}%
        \setcounter{equation}{0}
        \renewcommand{\theequation}{S\arabic{equation}}
         \setcounter{section}{0}
        \renewcommand{\thesection}{S\arabic{section}}
 }
 
\clearpage
\begin{widetext}

\section*{Appendix}

\subsection*{Covered area for other values of $\beta$}

We present in this section a similar plot to that of \Cref{fig:image5}---comparing the covered area and the cost function--- for other values of $\beta.$ As mentioned there, $S$ shows decreasing behaviors for which $t_P>t_\mathcal{L}$ holds true (see \Cref{fig:image8}).

\begin{figure}[!h]
    \centering
\begin{subfigure}[b]{\textwidth}
\includegraphics[width=0.98\textwidth]{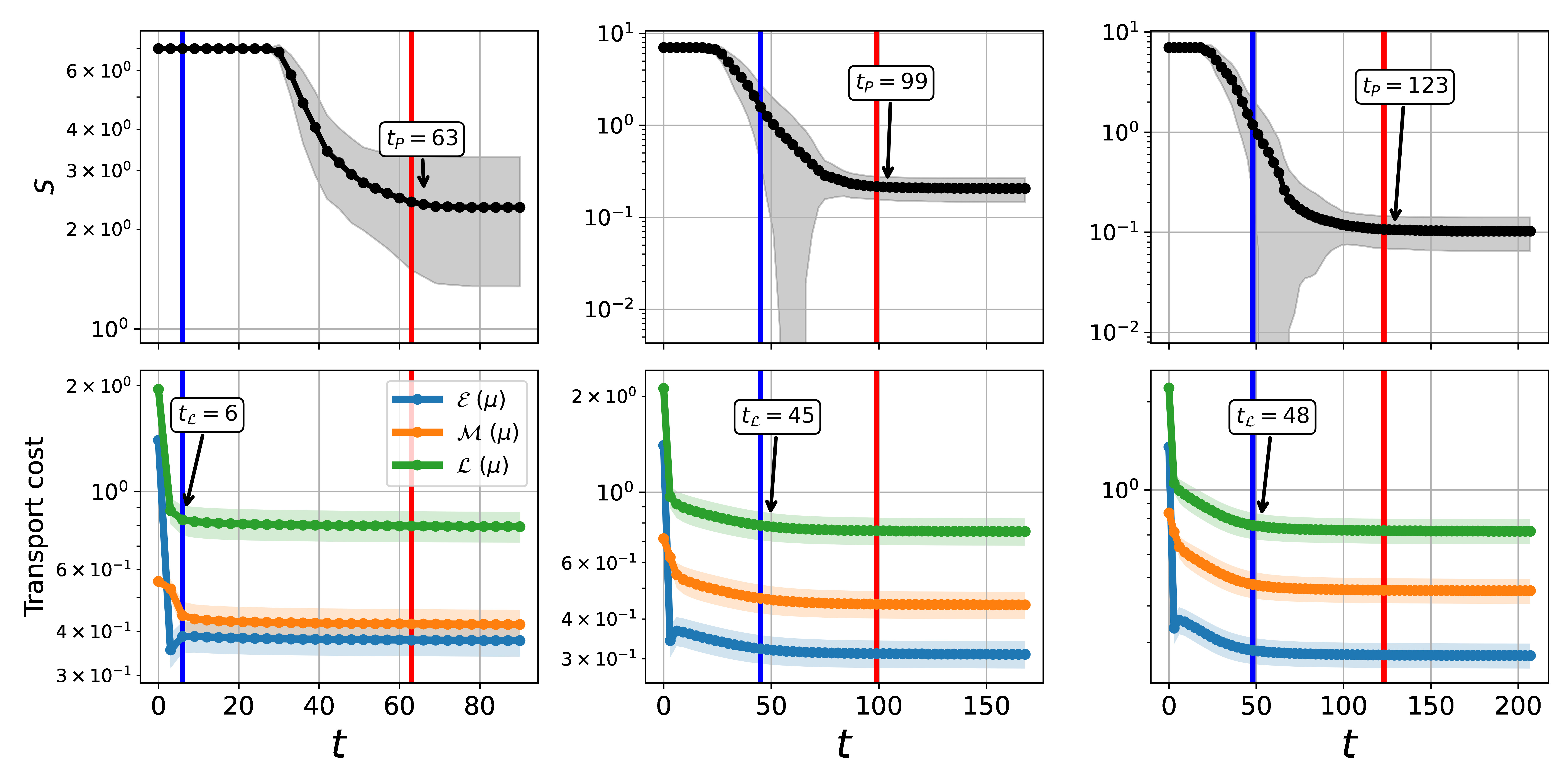}
\end{subfigure}
\begin{subfigure}[b]{\textwidth}
\includegraphics[width=0.98\textwidth]{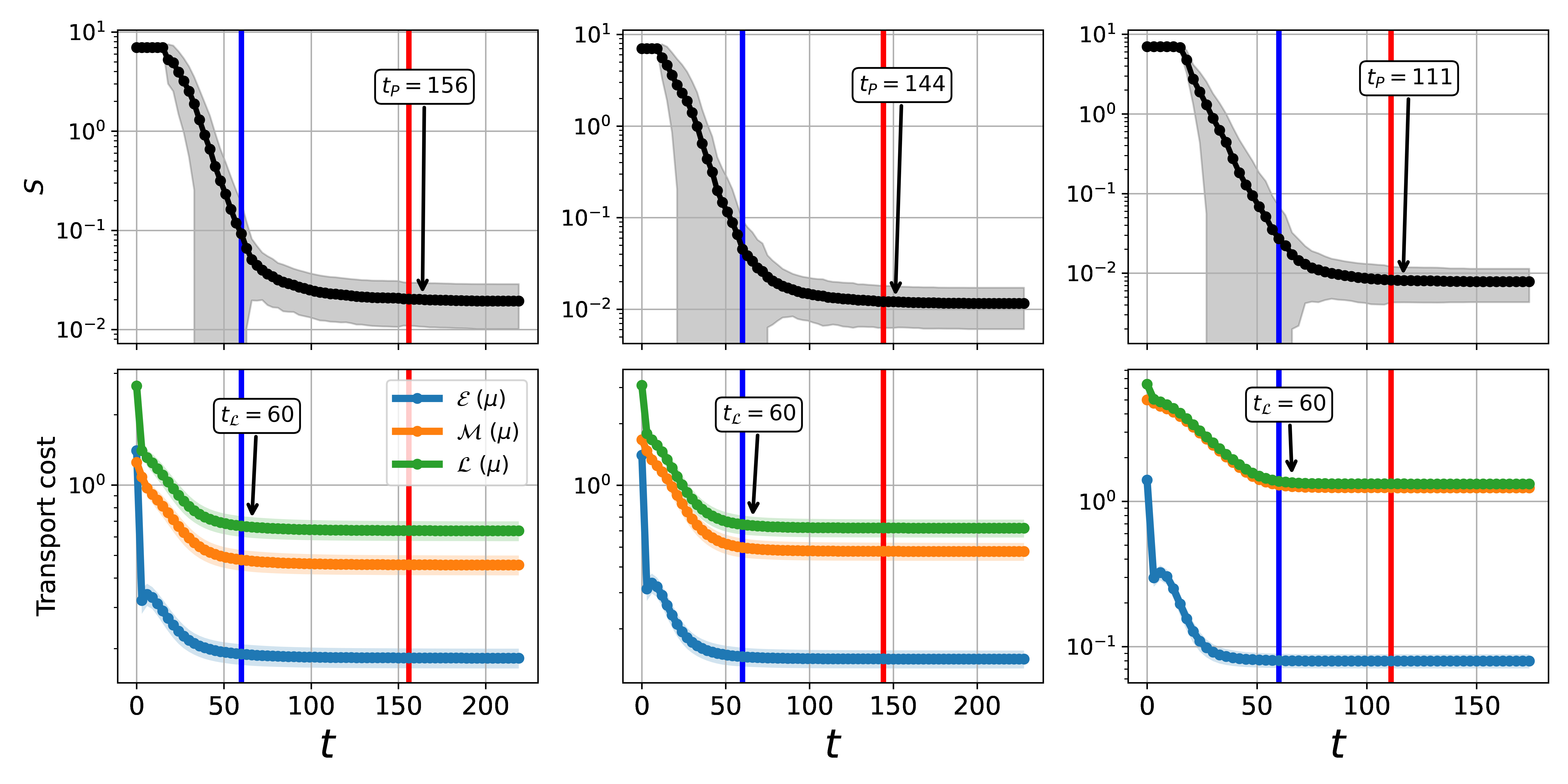}
\end{subfigure}
\caption{\textbf{$S$ and Lyapunov cost}. First and second top-down rows: from left to right we see $\beta=1.1, 1.3$ and $1.4$. Third  and fourth top-down rows: from left to right we see $\beta=1.6, 1.7$ and $1.9$. First and third top-down rows: mean and standard deviation of $S$ as a function of time $t$; Second and fourth top-down rows: Mean and standard deviation of  the Lyapunov cost $\mathcal{L}$, energy dissipation $\mathcal{E}$ and structural cost $\mathcal{M}$ of transport densities. Red and blue lines denote $t_P$ and $t_\mathcal{L}$ for $p = 1.05$.     }\label{fig:image8}
\end{figure}

\subsection*{Additional hypernetwork properties} 
In this section we extend the comparison between the cost function---minimized by the dynamics---and hypernetwork properties (see \Cref{fig:image9}). As mentioned in the main manuscript, similar monotonic behaviors can be observed in these cases.

\begin{figure}[!h]
    \centering
\includegraphics[width=0.98\textwidth]{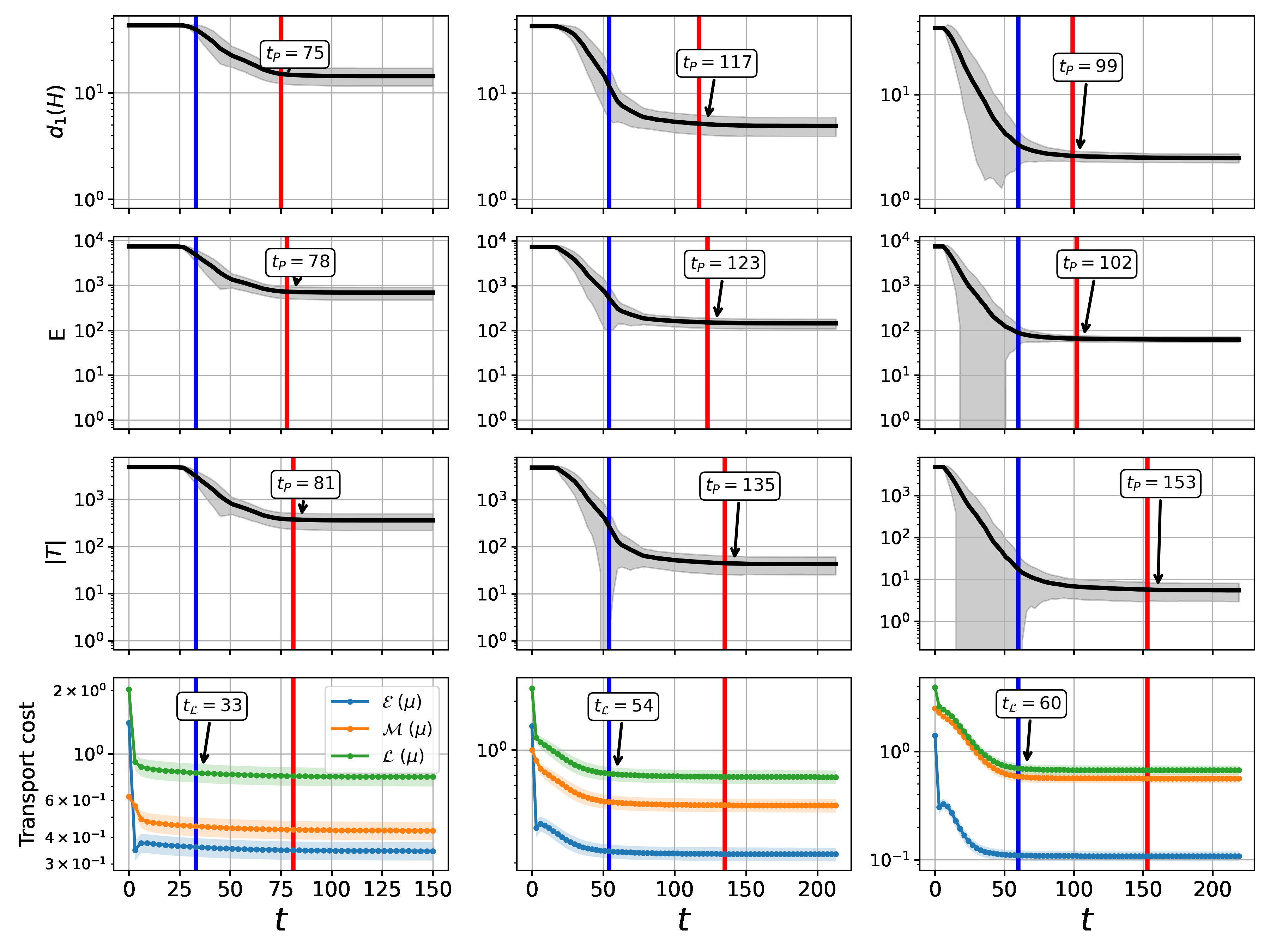}
\caption{\textbf{Other hypernetwork properties and Lyapunov cost}. From left to right: $\beta=1.2, 1.5$ and $1.8.$ From top to bottom: Mean and standard deviation of the average degree $d_1(H)$, number of hyperedges $|E_H|,$ number of triangles $|T|$, and the Lyapunov cost $\mathcal{L}$, energy dissipation $\mathcal{E}$ and structural cost $\mathcal{M}$. Red and blue lines denote $t_P$ and $t_\mathcal{L}$ for $p = 1.05$.}\label{fig:image9}
\end{figure}

\subsection*{\textit{P. polycephalum} hypernetworks}
\paragraph{Data information.} We explain in this section further details about the analyzed real data. \\
The images are taken from the Slime Mold Graph Repository \cite{dirnberger2017introducing} as mentioned in the main manuscript. We study  4  $\{H_i\}_i^T$ sequences of different lengths. The length ($T$) varies depending on the number of images included in the sequence. This is because different experiments need more o fewer shots. These experiments, as mentioned in the repository's documentation, consist of placing a slime mold inside a Petri dish with a thin sheet of agar where no food is provided. Slime mold's exploration of the dish, as explained by the creators, is unbiased, due to the lack of food.  Given that this organism is initially placed along one of the short edges of the rectangular dish, the experiment is considered to be finished once the plasmodium reaches the other short side. No more pictures are taken after this happens.
\paragraph{Hypergraph extraction.} We used the image sets \textit{motion12, motion24, motion40} and \textit{motion79}, located in the repository, to build the studied hypernetworks. These sets contain a number of images ranging from 60 to 150. Hypernetworks are then extracted using the \textit{Img2net} algorithm described in \cite{baptista2020principlednet} as mentioned in the main manuscript, using the same configuration described in \cite{baptista2021temporal}.

\begin{figure}[!h]
    \centering
\begin{subfigure}[b]{\textwidth}
\includegraphics[width=0.8\textwidth]{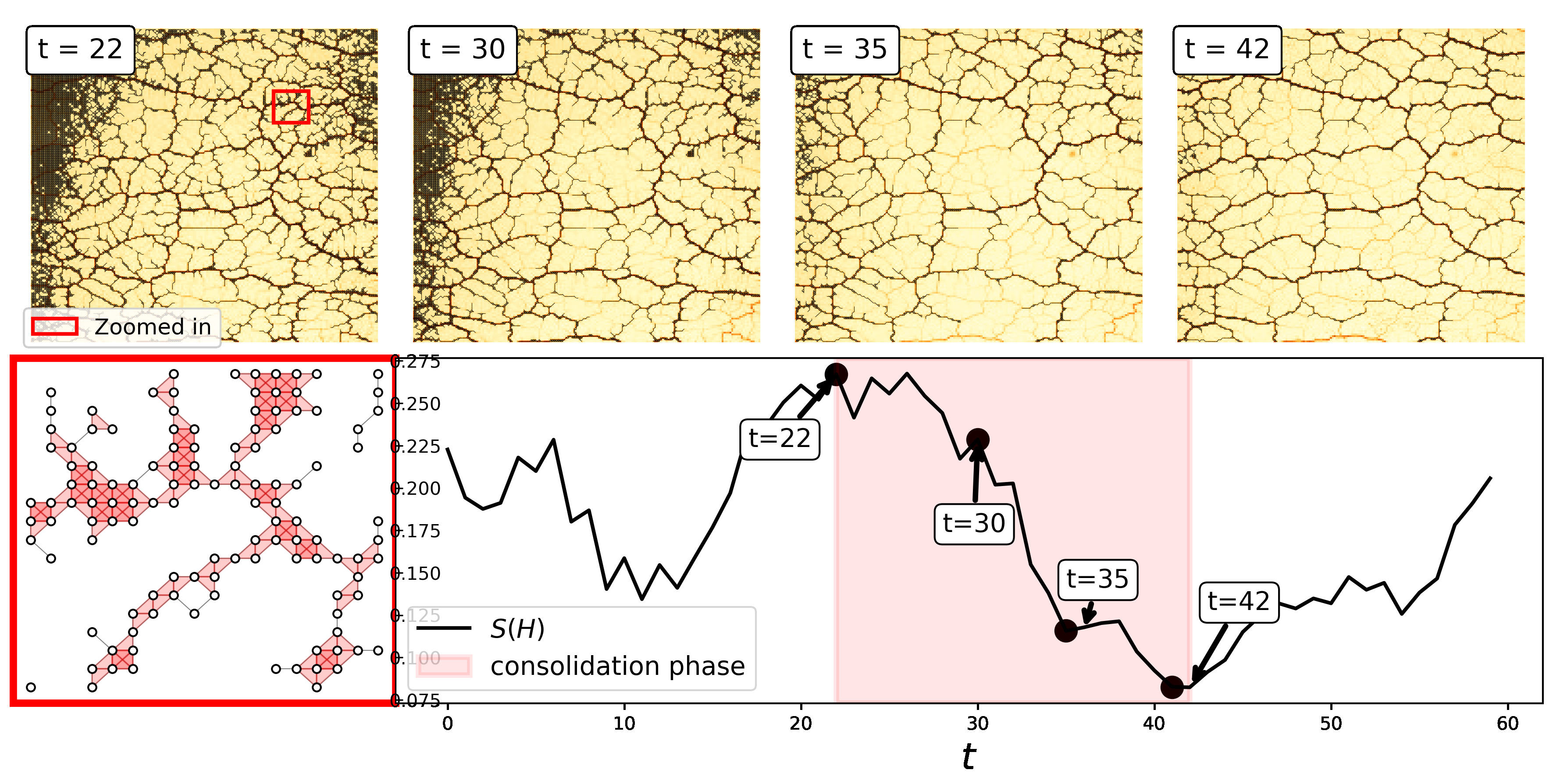}
\end{subfigure} 
\begin{subfigure}[b]{\textwidth}
\includegraphics[width=0.8\textwidth]{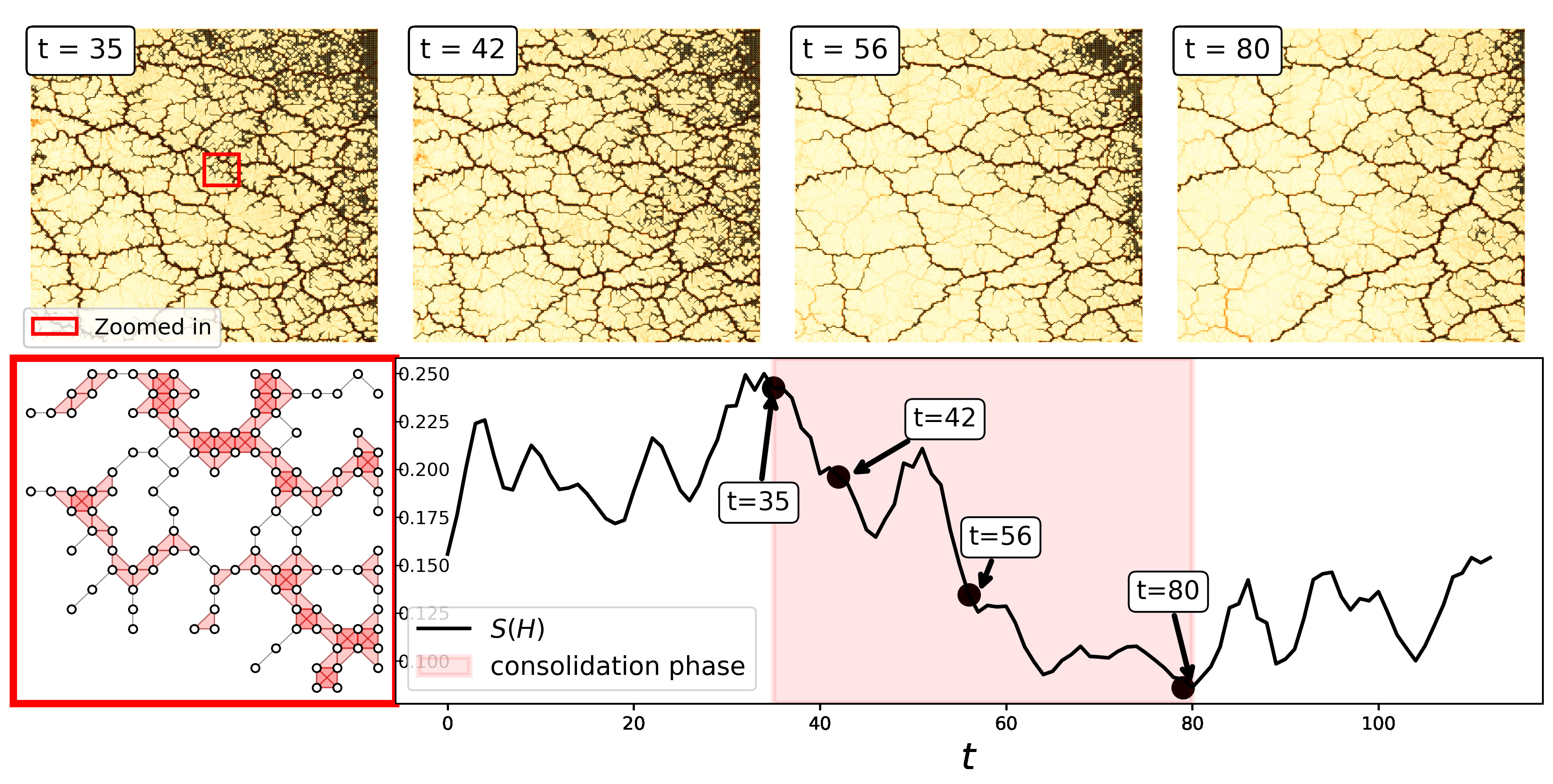}
\end{subfigure} 
\begin{subfigure}[b]{\textwidth}
\includegraphics[width=0.8\textwidth]{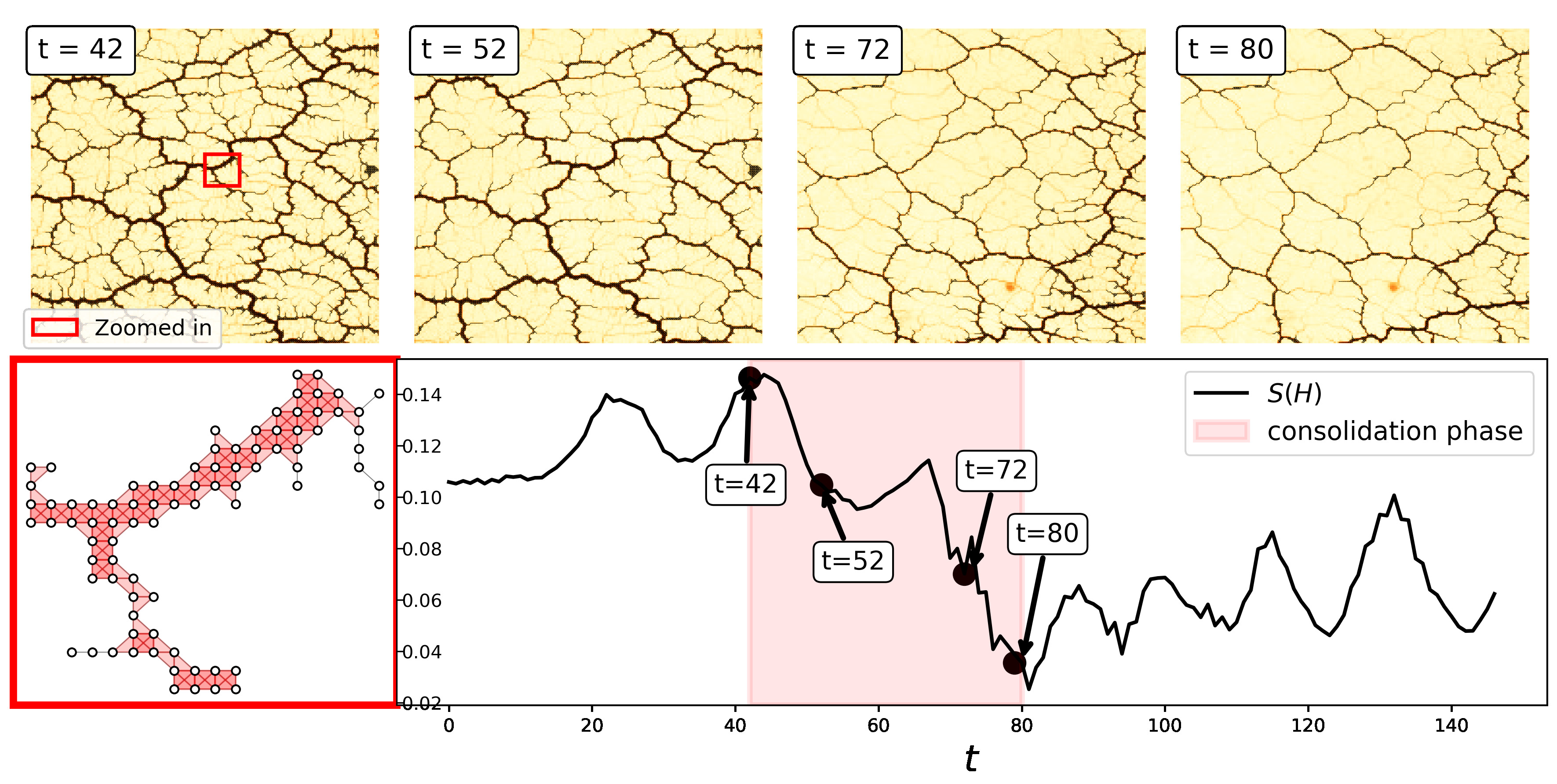}
\end{subfigure} 
\caption{\textbf{\textit{P. polycephalum} $S$ evolution}. From top to bottom: \textit{motion24, motion40} and \textit{motion79}. Plots are separated in couples. For every couple, the plots on top show both \textit{P. polycephalum} images and hypernetworks extracted from them. The network at the lower leftmost plot is a subsection of the hypergraph shown inside the red rectangle on top. The plot at the bottom shows the covered area as a function of time. The red shade in this plot highlights a tentative consolidation phase towards optimality.} \label{fig:image10}
\end{figure}
\end{widetext}

\end{document}